\documentclass[12pt]{iopart}

\usepackage{iopams}  
\usepackage{graphicx}  

\usepackage{xcolor}
\usepackage{lineno,hyperref}
\modulolinenumbers[5]

\newcommand{\up}{\uparrow}
\newcommand{\dn}{\downarrow}

\usepackage{harvard}
\begin{document}

\title[Hund's metal regimes and orbital selective Mott transitions]{Hund's metal regimes and orbital selective Mott transitions in three band systems}


\author{Jorge I. Facio$^{1,2}$, Pablo S. Cornaglia$^{1}$}

\address{$^1$Centro At{\'o}mico Bariloche and Instituto Balseiro, CNEA, CONICET, (8400) Bariloche, Argentina}
\address{$^2$Institute for Theoretical Solid State Physics, IFW Dresden, Helmholtzstr. 20, 01069 Dresden, Germany}
\ead{j.facio@ifw-dresden.de}

\begin{abstract}
We analyze the electronic properties of interacting crystal field split three band systems. 
Using a rotationally invariant slave boson approach we analyze the behavior of the electronic mass renormalization as a function of the intralevel repulsion $U$, the Hund's coupling $J$, the crystal field splitting, and the number of electrons per site $n$.
We first focus on the case in which two of the bands are identical and the levels of the third one are shifted by $\Delta>0$ with respect to the former. We find an increasing quasiparticle mass differentiation between the bands, for system away from half-filling ($n=3$), as the Hubbard interaction $U$ is increased. This leads to orbital selective Mott transitions where either the higher energy band (for $4>n>3$) or the lower energy degenerate bands ($2<n<3$) become insulating for $U$ larger than a critical interaction $U_{c}(n)$.
	Away from the half-filled case $|n-3|\gtrsim 0.3$ there is a wide range of parameters for $U<U_c(n)$ where the system presents a Hund's metal phase with the physics dominated by the local high spin multiplets.
Finally, we study the fate of the $n=2$ Hund's metal as the energy splitting between orbitals is increased for different possible crystal distortions. We find a strong sensitivity of the Hund's metal regime to crystal fields due to the opposing effects of $J$ and the crystal field splittings on the charge distribution between the bands.
\end{abstract}


\section{Introduction}

The importance of the Hund's rule coupling $J$ to enhance electronic correlations in metallic systems with Coulomb interactions well below the critical Mott transition values has been recognized in recent years \cite{haule2009,luca2011,georges2013,luca2017,STADLER2018}.
Systems in this regime have been dubbed ``Hund's metals'' and notable examples that seem to fit into this class can be found in different material families, like Ru based oxides \cite{mravlje2011coherence} and Fe based superconductors \cite{haule2009,PhysRevLett.121.187003}.
While these families include systems in which the transition metal atom has very different crystal environments, a shared property is that due to the oxidation state (Ru$^{4+}$ or Fe$^{2+}$), the electronic density in the relevant low-energy manifold of electronic bands is different from half-filling.
This is an essential characteristic of Hund's metals since in such conditions $J$ acts to significantly increase the critical Coulomb interaction of the metal to Mott insulator transition. 

Crystal field splittings compete with the Hund's coupling inducing charge imbalances between the different orbitals that can produce a differentiation of the quasiparticle masses between the associated electron bands, even leading to orbital selective Mott transitions (OSMTs) in which only a subset of the bands becomes insulating.  
Fe$^{2+}$ based materials experimentally found to be Mott insulators like La$_2$O$_3$Fe$_2$Se$_2$ \cite{PhysRevLett.104.216405} and BaFe$_2$S$_3$ \cite{PhysRevLett.115.246402,takahashi2015pressure,takubo2017orbital,materne2018bandwidth}, have been theoretically predicted to present OSMTs upon doping or application of hydrostatic pressure \cite{Capone2015HundCF,craco2018microscopic,patel2018orbital}.
In ruthenates, the interest in OSMTs was sparked by the intriguing experimental observations \cite{PhysRevLett.84.2666,PhysRevB.62.6458} and the prediction of coexistence of metallic and insulating behavior in Sr$_{2-x}$Ca$_x$RuO$_4$ \cite{anisimov2002orbital}.
OSMTs have since been predicted to occur in a variety of physical situations including systems with asymmetric bands \cite{anisimov2002orbital,Ruegg2005} (different bandwidths or densities of states in the non interacting limit), crystal field split bands,  \cite{de2005orbital,Vojta2010,song2015possible}, and even in momentum space on single band systems \cite{ferrero2009a,ferrero2009}.

We revisit this problem analyzing in a three band model the role of the different competing energy scales, including crystal field splittings that lift the band degeneracy, for arbitrary orbital occupation. 
Three bands models are expected to capture the essential physics of $t_{2g}$ shells in transition metal oxides as the ruthenates, vanadates, and titanates \cite{werner2009} and to serve as a guide for understanding systems with a higher number of relevant orbitals~\cite{STADLER2018}. 
Our main focus is the behavior of the electron correlations, as measured by the quasiparticle mass, the stability of the Hund's metal regime and the presence of OSMTs. 
To that aim we perform rotationally invariant slave-boson mean field theory (RISB) calculations \cite{Lechermann2007,ferrero2009a,ferrero2009}, within the single-site dynamical mean field theory (DMFT) approximation\cite{georges1992hubbard,georges1996dynamical}. 
The RISB method has proven to be a fast and reliable impurity solver for DMFT equations in the metallic phase, being able to capture Hund's metal physics \cite{facio2017nature,PhysRevB.97.125154} and to describe OSMTs \cite{ferrero2009,ferrero2009a}.

The rest of this article is organized as follows. 
In Section \ref{methods} we present the Hamiltonian analized as well as a benchmark in the absence of crystal fields between RISB and CTQMC.
In Section \ref{tetra_1} and \ref{tetra_2} we analyze for arbitrary electronic densities the electronic correlations in the presence of a distortion that shifts the energy of one orbital while keeping the others degenerated.
In Section  \ref{ortho} we analyze the evolution of the $n=2$ Hund's metal under the different possible distortions for a three-orbital system. 
Finally, in Section \ref{summary} we present our conclusions.

\section{Methods}
\label{methods}
We consider the Slater-Kanamori Hamiltonian for a three band system
\begin{eqnarray}
	H_K &=&\sum_{i,j,m,m^\prime,\sigma}t_{ij}^{mm^\prime}d^\dagger_{im\sigma}d^{}_{jm^\prime\sigma}+\sum_{i,m,\sigma}(\varepsilon_{m}-\mu)n_{im\sigma}\nonumber\\ 
	&+& U\sum_{i,m} n_{im\up} n_{im\dn} + U' \sum_{im\neq m'} n_{im\up} n_{im'\dn}\nonumber\\
	& +& (U'-J) \sum_{i,m < m',\sigma} n_{im\sigma} n_{im'\sigma} \\ 
	&-& J\sum_{i,m\neq m'} d^\dag_{im\up} d_{im\dn} d^\dag_{im'\dn} d_{im'\up} \nonumber\\
	&+& J \sum_{i,m\neq m'} d^\dag_{im\up} d^\dag_{im\dn} d_{im'\dn} d_{im'\up}\nonumber,
\end{eqnarray}
where $t_{ij}^{mm^\prime}$ is a hopping term between orbital $m$ on site $i$ and orbital $m^\prime$ on site $j$, $\varepsilon_m$ is the crystal field energy, $\mu$ is the chemical potential, 
$J$ is the Hund's rule coupling, and $U$ and $U^\prime$ are the intraorbital and interorbital interactions, respectively. 
We consider, for simplicity, a semicircular density of states for each orbital:
\begin{equation}
        D(\varepsilon)=\frac{2}{\pi D}\sqrt{1-(\varepsilon/D)^2},
        \label{eq:ldos}
\end{equation}
where $D$ is the half-bandwidth of the conduction electron band in the absence of interactions.
The interlevel interaction $U^\prime$ in a rotational symmetry (atom in free space) reads $U^\prime=U-2 J$, which as expected satisfies $U^\prime<U$. As it is customarily done, we will use this expression in what follows. 

\begin{figure}[t]\center
\includegraphics[width=0.5\columnwidth,angle=0,keepaspectratio=true]{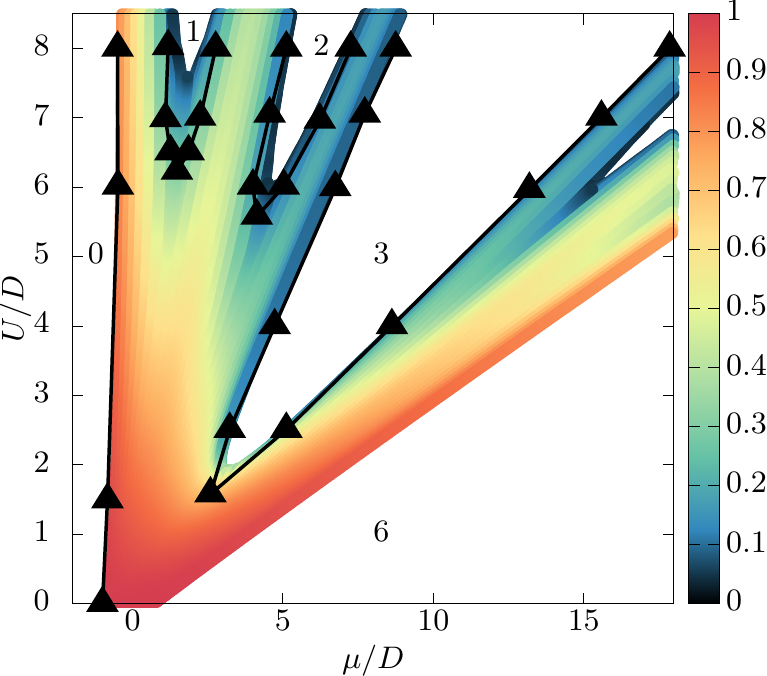}
	\caption{Zero crystal field splitting phase diagram in the chemical potential $\mu$ vs. local interaction $U$ space. 
	The black triangles correspond to DMFT results using CT-QMC, taken from Ref. \protect\cite{werner2009}. 
	The quasiparticle weight as obtained from RISB is presented using a color scale. The Hund's coupling is $J=0.167 U$.}
\label{fig:phased}
\end{figure}

We use the RISB method to solve the DMFT equations and obtain the mass renormalization 
\begin{equation}\label{eq:qp}
	Z_\alpha=\frac{m\,\, }{m^\star}=\left(1-\left. \frac{\partial \Sigma_\alpha(\omega)}{\partial \omega}\right|_{\omega=0}\right)^{-1},
\end{equation}
where $\Sigma_\alpha(\omega)$ is the local self energy for orbital $\alpha$. 

A detailed analysis of this model using Dynamical Mean Field Theory (DMFT) with continuous time quantum Monte Carlo as the impurity solver was presented by Werner {\it et al.} in Ref. \cite{werner2009} but no information on the quasiparticle renormalization was provided there. In Fig. \ref{fig:phased} we present the phase diagram obtained by Werner {\it et al.} in the abscence of crystal fields (black triangles) together with the calculated quasiparticle weight as obtained by the RISB method. We find a good agreement for the Mott transition lines although RISB overestimates the critical interaction for $n=1$ by $\sim 20\%$. 

\section{Results}

We analyze first the effects of a tetragonal distortion which leads to $\varepsilon_1=\varepsilon_2=0$ and $\varepsilon_3=\Delta$.
In Section \ref{tetra_1} we study the half-filled case while in Section \ref{tetra_2} we analyze arbitrary electronic densities.
Last, in Section \ref{ortho} we study the $n=2$ Hund's metal under different crystals distortions.

\subsection{Half-filled}
\label{tetra_1}
\begin{figure}[tb]\center
\includegraphics[width=0.5\columnwidth,angle=0,keepaspectratio=true]{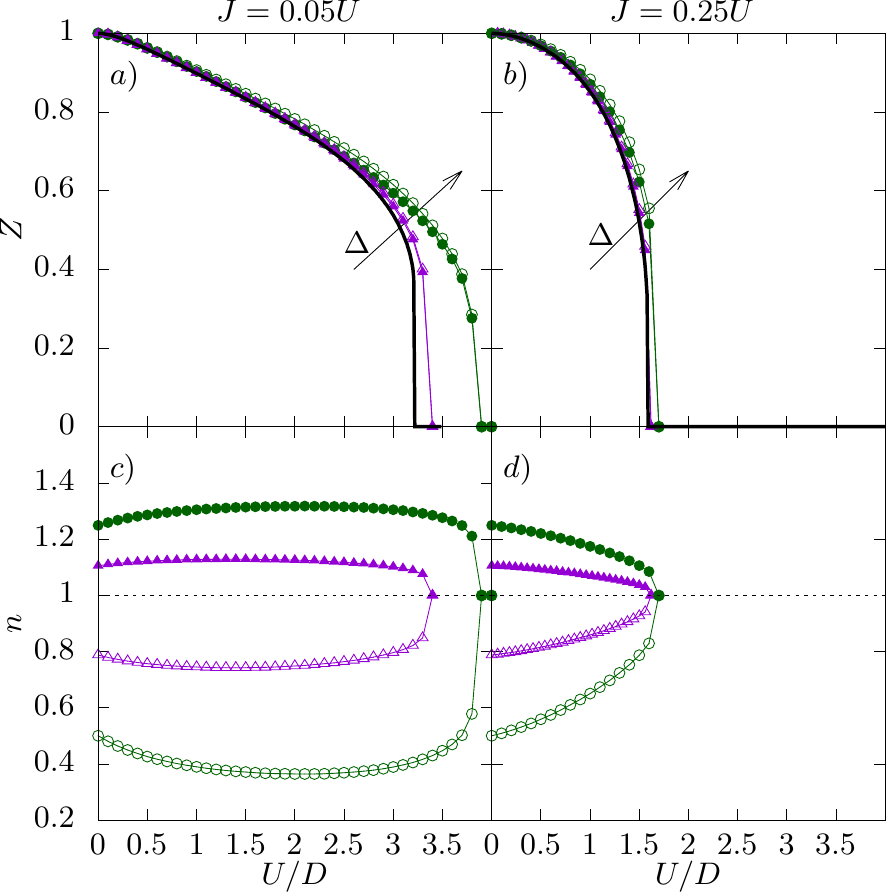}
	\caption{Quasiparticle mass enhancement $Z$ (top panels) and orbital occupancy $n$ (lower panels), as a function of the local interaction $U/D$ in a half-filled system (three electrons per site). 
	Open symbols correspond to orbital $3$ while filled symbols to orbitals $1$ and $2$.  
	The value of $J/U$ for the left and right panels is indicated in the figure. The crystal field splitting $\Delta$ is 0 (solid lines), 0.25$D$ (triangles), and 0.6$D$ (circles). }

\label{halffilled}
\end{figure}
We first consider the half-filled case with an orbital occupancy of 3 electrons per site.
In the $\Delta=0$ case the three orbitals have the same occupancy and the main effect of $J>0$ is to shift the critical interaction $U_{c}$ to lower values as the charge excitation gap increases to $U+2J$ \cite{haule2009,luca2011,luca2017,isidori2018charge}. The transition is of the first order type with a finite jump in $Z$ at the transition (see Fig. \ref{halffilled})\cite{facio2017nature}. A finite $J$ breaks the degeneracy of the atomic ground state favoring a $S=3/2$ configuration with an occupancy of a single electron per orbital.
The crystal field splitting $\Delta>0$ makes it energetically unfavorable to occupy the higher energy orbital and produces (in the metallic phase) a charge transfer between the orbitals. The charge redistribution between orbitals is dominated by the ratio $\Delta/J$. The crystal field splitting competes with $J$ that favors an even distribution of the charge between the orbitals. This competition manifests itself in the behavior of the critical interaction $U_{c}$ with $J$ and $\Delta$. While increasing $J$ leads to a larger charge excitation gap and to a reduction of $U_{c}$, increasing $\Delta$ for a fixed $J/U$ produces an enhancement of $U_{c}$. A non-zero $\Delta$ does not change the nature of the transition but makes it concomitant with a sudden charge redistribution such that each orbital has an occupancy of one electron in the insulating phase.
Although the charging of the different orbitals can be quite different close to the Mott transition, the quasiparticle weight does not show a significant orbital differentiation for the wide range of parameters analyzed.

\subsection{Doped system}

Ref. \cite{huang2012complete} analyzed the site occupation $n=4$ and found a rich phase diagram as a function of the level splitting and the Coulomb interaction $U$, including Mott and orbital selective Mott phases. We extend these results  to general doping levels and analyze the effect of the crystal field on the Hund's metal phase.

\label{tetra_2}
\begin{figure}[tb]\center
\includegraphics[width=0.5\columnwidth,angle=0,keepaspectratio=true]{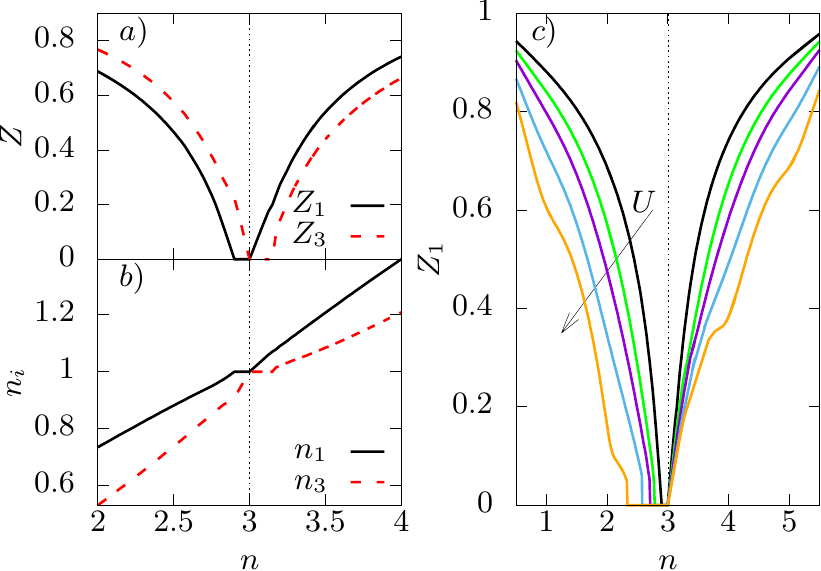}
	\caption{ $a)$ Quasiparticle weights  and b) orbital occupancy as a function of the site occupancy for $U=1.86 D$.
$c)$ Quasiparticle weight of the orbitals $1$ and $2$ for different values of $U/D=1.86,\,2.4,\,3.0,\,4.2$ y 6.0. The other parameters are $J=0.25 U$ and $\Delta=0.25D$.
}
\label{dopado}
\end{figure}
To stabilize an orbital selective Mott phase we may introduce a hole or electron doping to the system. As it was shown for a two orbital system with nonzero Hund's coupling and crystal field splitting \cite{werner2007high}, for a range of doping values, the charge is incorporated to one of the orbitals making it metallic while the other remain insulating. This behavior is also observed for a three orbital system as it is shown in Fig. \ref{dopado}. The parameters are such that the system is in the Mott phase for the undoped ($n=3$) case ($U=1.86D$, $J=0.25 U$, and $\Delta=0.25$, see Fig. \ref{halffilled}).
Electron doping increases the charge in orbitals 1 and 2 which become metallic, while the charge and the insulating nature of orbital 3 remain unchanged for a finite range of dopings, as it can be seen form the behavior of the corresponding quasiparticle weights [see Fig. \ref{dopado}a)]. On the contrary, upon hole doping orbital 3 decreases its charge and becomes metallic while orbitals 1 and 2 remain insulating with an occupancy $n_1=n_2=1$. For large enough electron or hole doping, the three orbitals become metallic through an orbital selective Mott transition. 
As it can be seen in Fig. \ref{dopado}c), the range of doping in which the orbital selective Mott phase is observed increases with increasing $U$.

The critical interaction for the OSMT is dominated by the occupancy of the different orbitals and at least one of the orbitals must have an integer occupancy in the OSMP. 
As in the half-filling case, the charge distribution between the orbitals depends on $\Delta$ and $J$. We analyze below the role of these parameters on the OSMT. 
\begin{figure}[tb]\center
\includegraphics[width=0.5\columnwidth,angle=0,keepaspectratio=true]{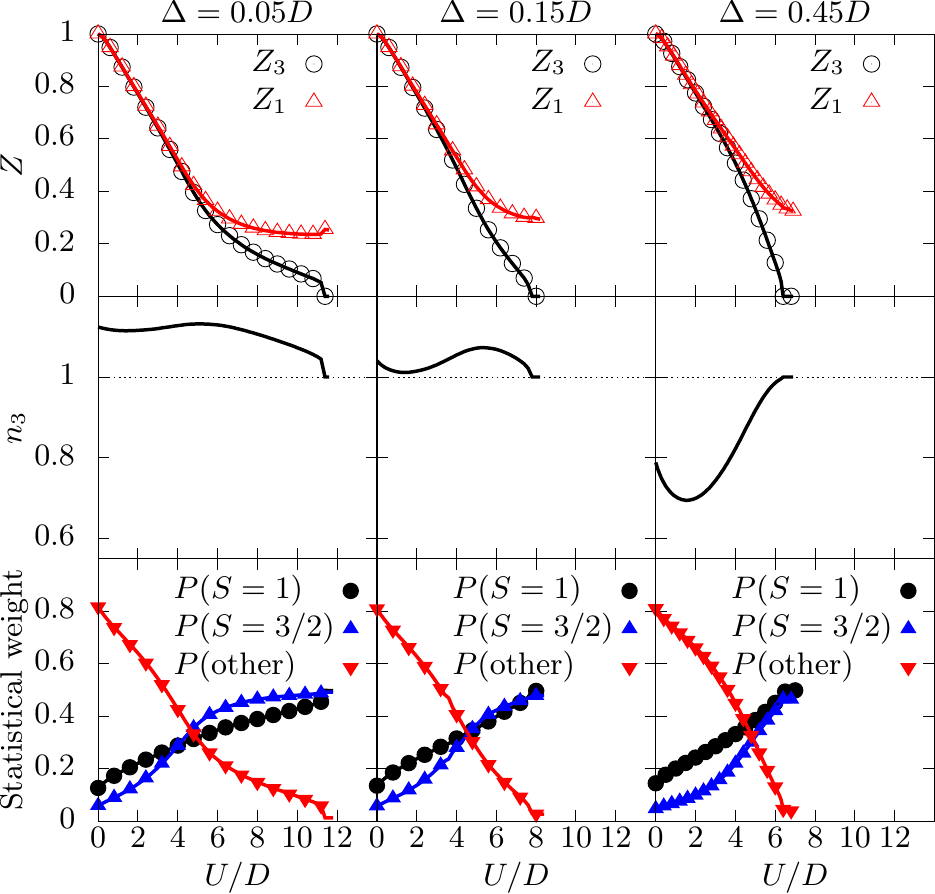}
\caption{\textit{Upper panels:} quasiparticle weight as a function of $U/D$ for different crystal field splittings $\Delta$ indicated in the figure, $J/U=0.05$, and $n=3.5$.
\textit{Middle panels:} orbital 3 occupancy $n_3$ vs. $U/D$.
\textit{Lower panels:} statistical weight of different atomic multiplets vs. $U/D$.
}
\label{z_orb}
\end{figure}
We first focus on the electron doped case with an occupancy of $3.5$ electrons per site. In this case, for $J\neq 0$, orbital 3 becomes insulating for $U>U_c$ while orbitals 1 and 2 remain metallic. In Fig. \ref{z_orb} we present the quasiparticle weight and the occupancy of orbital 3 as a function of $U$. For the values of $\Delta<2D$ presented in the figure, there is a partial occupancy of orbital 3 even in the non-interacting limit as the bandwidth is larger than the energy shift between the band associated with orbital 3 and the bands associated with orbitals 1 and 2. As the interaction is increased the quasiparticle bandwidths ($Z_\alpha D$) decrease and $\Delta$ becomes more effective polarizing the charge. An increasing $U$ also leads to an increase in $J=0.05U$ which favors an even distribution of the charge between orbitals. For $U\sim 0$ ($J\ll \Delta$) the former effect dominates the physics and $n_3$ decreases. For larger values of $U$ we obtain two different behaviors depending on the value of $\Delta$. For $\Delta \ll D $ there is a wide range of interaction parameters where the system presents a Hund's metal behavior with the physics dominated by the $S=3/2$ and $S=1$ multiplets (see lower panels in Fig. \ref{z_orb}). This regime is marked by an increasing effective mass differentiation between the orbitals with an increasing interaction $U$. For larger values of $\Delta$, the range of parameters where the system is in this regime is reduced as the critical interaction for the OSMT decreases. 
\begin{figure}[tb]\center
\includegraphics[width=0.5\columnwidth,angle=0,keepaspectratio=true]{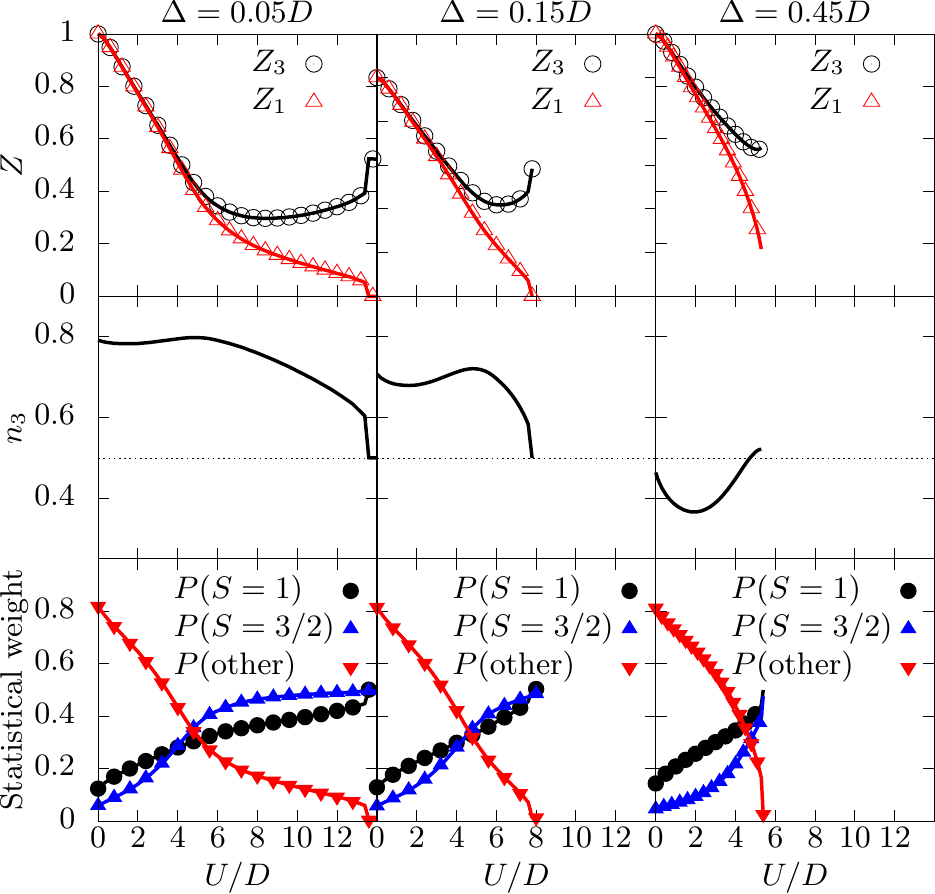}
\caption{\textit{Upper panels:} quasiparticle weight as a function of $U/D$ for different crystal field splittings $\Delta$ indicated in the figure, $J/U=0.05$, and $n=2.5$.
\textit{Middle panels:} orbital 1 occupancy $n_1$ vs. $U/D$.
\textit{Lower panels:}  statistical weight of different atomic multiplets vs. $U/D$. 
}
\label{two_lj}
\end{figure}

For hole doping, orbitals 1 and 2 become insulating at a critical interaction, while orbital 3 remains metallic. This is illustrated in Fig. \ref{two_lj} for an occupancy $n=2.5$ and the other parameters as in Fig. \ref{z_orb}. The overall behavior as a function of $U$ is similar to the $n=3.5$ case, the main difference being that the behavior of the quasiparticle masses is interchanged with $Z_1=Z_2$ lower than $Z_3$ and vanishing at the OSMT.

\begin{figure}[tb]\center
\includegraphics[width=0.4\columnwidth,angle=0,keepaspectratio=true]{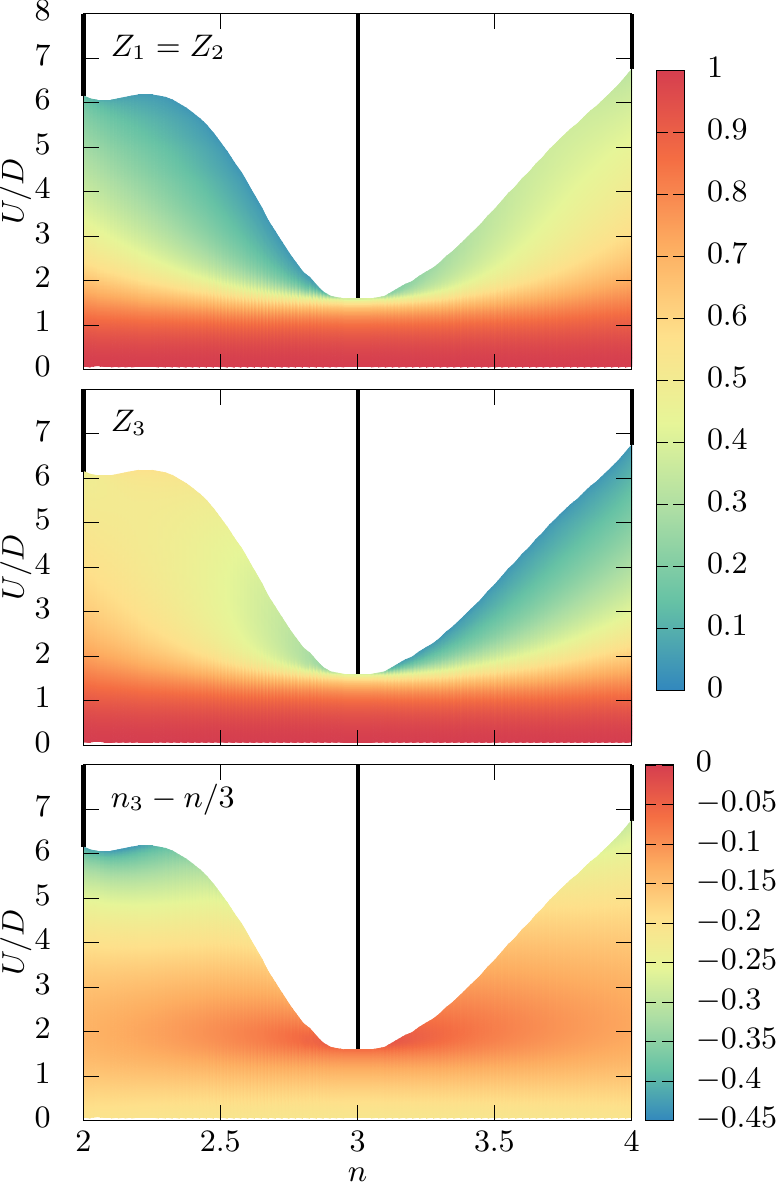}
	\caption{\textit{Top and middle panels:} quasiparticle weights as function of $U/D$ and $n$. \textit{Lower panel:} Ocupancy of orbital 3 relative to its value at zero crystal field. The white areas correspond to parameters where at least one of the bands is insulating. The other parameters are $J/U=0.25$, $\beta D = 400$, and $\Delta/D=0.25$.
}
\label{semi}
\end{figure}
The main results of this section are presented in Fig. \ref{semi} which shows the quasiparticle weight for orbital 3 and orbitals 1 and 2 in the regions of the $n$ vs $U/D$ plane where the system is in the metallic phase. The white area corresponds to an OSMP where only one of the bands is insulating ($n<3$), two bands are insulating ($n>3$) or to a Mott insulator where the three bands are insulating (integer fillings $n=2$, $n=3$, and $n=4$). Away from the $|n-3|\lesssim 0.3$ cases the system presents a wide range of Coulomb interactions where one (or two) of the bands is strongly correlated. This regime correspond to the Hund's metal phase where the physics is dominated by the high spin multiplets $S=3/2$ and $S=1$.

	\begin{figure}[tb]\center
\includegraphics[width=0.5\columnwidth,angle=0,keepaspectratio=true]{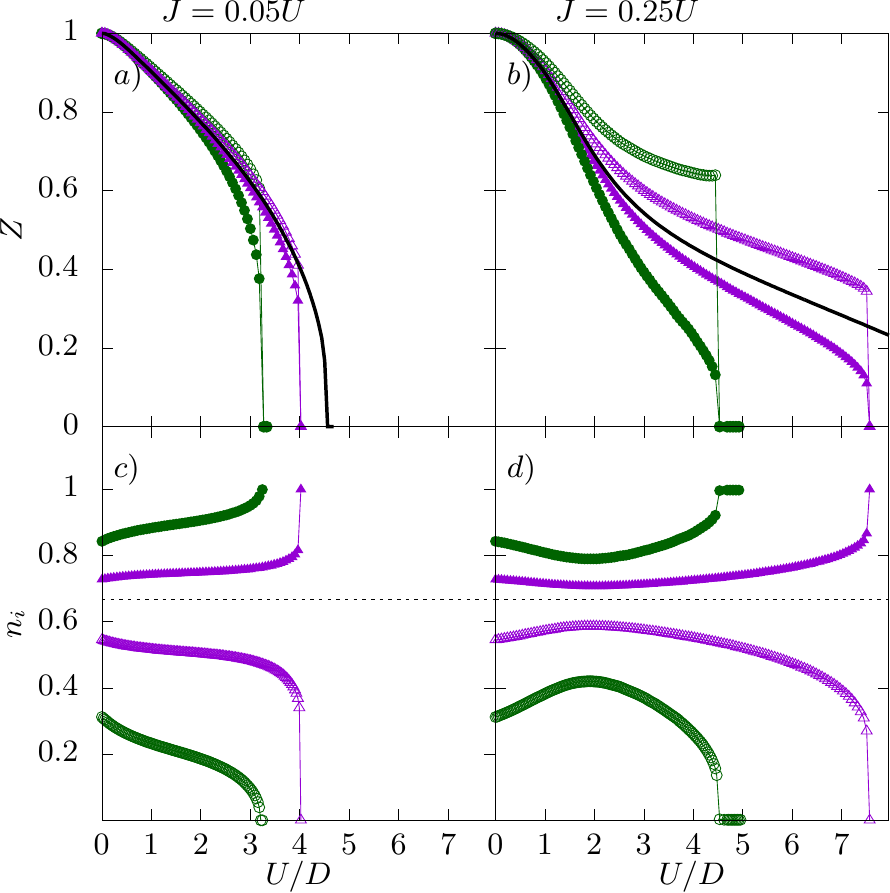}
\caption{
	Quasiparticle mass enhancement $Z$ (top panels) and orbital occupancy $n$ (lower panels), as a function of the local interaction $U/D$ for a two electrons per site occupancy ($n=2$). 
	Open symbols correspond to orbital $3$ while filled symbols to orbitals $1$ and $2$.  
	The value of $J/U$ for the left and right panels is indicated in the figure. The crystal field splitting $\Delta$ is 0 (solid lines), 0.15 (triangles), and 0.45 (circles).
	}
\label{cfn2j}
\end{figure}

\subsection{Hund's metal vs crystal field distortions}
\label{ortho}
Here we analyze how the Hund's metal phase evolves as different crystal field distortions are enabled. 
We focus on the case $n=2$ and consider the three possible distortions in a three-band model: one orbital energy pushed above or below the energy of a double-degenerated set of orbitals or the three orbitals split in energy. In the context of $t_{2g}$ shells of transition metal oxides, the first two cases correspond to a tetragonal distortion whereas the third one to the orthorhombic case.

	In Fig. \ref{cfn2j}~we present the quasiparticle mass enhancement and the orbital occupancies for a three orbital system with a tetragonal distortion and two values of the $J/U$ ratio. For low $J$ values ($J/U=0.05$), increasing the crystal field splitting $\Delta$ leads to a decrease in the critical interaction and to an increase in the mass enhancement differentiation.  In the high $J/U=0.25$ regime, the system has a much larger critical interaction $U_c(\Delta=0)$ in the $\Delta=0$ case and presents a wide range of Coulomb interactions $U$ where the quasiparticle mass is strongly renormalized even though $U \ll U_c(\Delta=0)$. In this regime the system is strongly sensitive to changes in the crystal field splitting which leads to a strong orbital differentiation in the quasiparticle mass enhancement and to a significant reduction of the critical Coulomb interaction.
	
\begin{figure}[tb]\center
\includegraphics[width=8.5cm,angle=0,keepaspectratio=true]{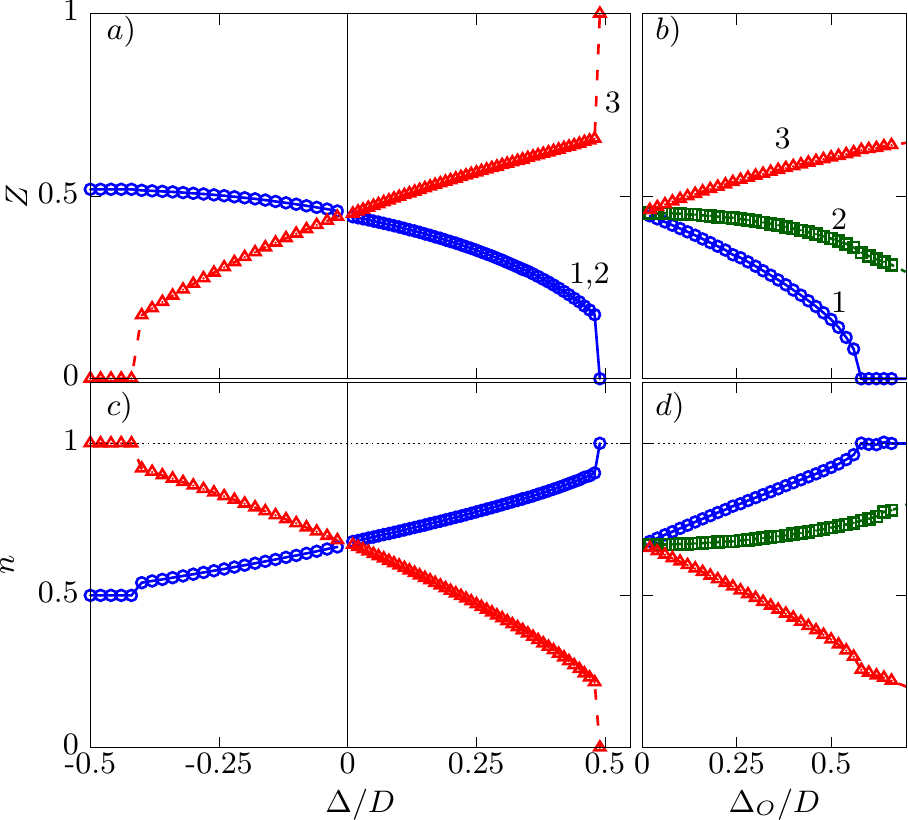}
\caption{\textit{Upper panels:} Quasiparticle weight as function of crystal field for different distortions and interaction parameters $U=4D=4J$. $a)$ Tetragonal case with level energies $\varepsilon_i = (0,0,\Delta)$. $b)$ Orthorrombic case with $\varepsilon_i = (0,\Delta_O/2,\Delta_O)$. 
\textit{Lower panels:} Orbital occupancies for each case.
}
\label{n2}
\end{figure}

	We now fix the interaction parameters to $U=4J=4D$, values which in the absence of crystal fields place the system in the Hund's metal phase, and study the evolution of the system for different possible crystalline distortions presented above. 
Fig. \ref{n2}$a-b)$ shows for the different distortions the quasiparticle weight as a function of crystal field.
In the three cases, as the energy splitting between orbitals is increased the quasiparticle weights differentiate becoming smaller for the orbitals whose occupancy approach 1.
In the tetragonal case having as lowest energy orbitals the double-degenerated set (positive $\Delta$ in Fig. \ref{n2}$a)$, as $\Delta$ is increased the Hund's metal evolves to a Mott insulator in which the third orbital has a vanishing occupancy. 
For the two other distortions the lowest energy orbital has only the spin-degeneracy and the Hund's metal evolves first to an OSMP in which the lowest energy orbital becomes insulating while the other two remain metallic. 

\section{Summary and conclusions}
	\label{summary}
	In summary, we have analyzed the electronic properties of a three band Hubbard system under a crystal field that partially lifts the orbital degeneracies. We focused on the low energy properties and characterized the electronic correlations through the quasiparticle mass renormalization $Z_\alpha$ for each band.
	Away from half-filling ($|n-3|\gtrsim 0.3$) the system presents Hund's metal behavior, with the physics dominated by local high spin multiplets, for a wide range of interaction parameters.

	We found a strong sensitivity of the Hund's metal to crystal fields that lift the degeneracies of the bands. In this regime, the crystal fields produce a strong quasiparticle mass differentiation between the bands and reduce significantly the range of parameters in which the Hund's metal regime is stable driving the system into a OSMTP.
 
\section*{Acknowledgements}
	We thank Gustavo Berman and Ulrike Nitzsche for technical assistance.
\bibliography{references}

\end{document}